\documentclass[twocolumn,aip,apl,reprint,amsmath,amssymb]{revtex4-1}
\usepackage[hidelinks]{hyperref}
\usepackage{graphicx,color,siunitx,bm}

\begin{document}

\title{Two-dimensional InSe as a potential thermoelectric material}

\author{Nguyen T. Hung}
\email{nguyen@flex.phys.tohoku.ac.jp}

\author{Ahmad R. T. Nugraha}
\email{nugraha@flex.phys.tohoku.ac.jp}

\author{Riichiro Saito}

\affiliation{Department of Physics, Tohoku University, Sendai
  980-8578, Japan}


\begin{abstract}
  Thermoelectric properties of monolayer indium selenide (InSe) are
  investigated by using Boltzman transport theory and first-principles
  calculations as a function of Fermi energy and crystal orientation.
  We find that the maximum power factor of p-type (n-type) monolayer
  InSe can be as large as 0.049 (0.043) W/K$^2$m at 300 K in the
  armchair direction.  The excellent thermoelectric performance of
  monolayer InSe is attributed to both of its Seebeck coefficient and
  electrical conductivity.  The large Seebeck coefficient originates
  from the moderate (about 2 eV) band gap of monolayer InSe as an
  indirect gap semiconductor, while its large electrical conductivity
  is due to its unique two-dimensional density of states (DOS), which
  consists of an almost constant DOS near the conduction band bottom
  and a sharp peak near the valence band top.
\end{abstract}

\date{\today}
\maketitle

Recent advances in the fabrication and characterization of
two-dimensional (2D) materials such as the transition metal
dichalcogenides (TMDs), black phosphorus (BP), and group III
chalcogenides have allowed researchers to look up unique electronic
properties of the materials and utilize them in various electronic
applications.~\cite{wang12-review2D,butler13-review2D,geim13-vdwreview}
Research on thermoelectricity, which is intended to convert waste heat
into electric energy, should also benefit from the advances of the 2D
materials.  Unlike graphene, the TMDs, BP, and group III chalcogenides
have sizable band gaps that could enable the enhancement of
thermoelectric properties due to the quantum confinement effects in
the low-dimensional
semiconductors.~\cite{dresselhaus2007new,hung2016quantum} It is thus
important to predict the best thermoelectric 2D material
theoretically.

A good thermoelectric material is characterized by how efficient
electricity can be obtained for a given heat source, in which two
quantities are often used for evaluation: (1) power factor,
$\text{PF}=S^2\sigma$, where $S$ is the Seebeck coefficient and
$\sigma$ is the electrical conductivity; and (2) figure of merit
$ZT=S^2\sigma\kappa^{-1}T$, where $\kappa$ is the thermal conductivity
and $T$ is the average absolute temperature.  The PF specifies how
much electricity can be generated, while $ZT$ specifies how efficient
electricity is obtained for a given temperature difference.  The
improvement of thermoelectric devices thus strongly depends on the
optimization of electronic and thermal transport properties, in which
the 2D materials may serve as a good
candidate.~\cite{saito2016gate,fei2014enhanced,bolin15-elph,yoshida2016gate,morteza-mos2}
For example, using the electric-double-layer transistor configuration,
it was found that the Seebeck coefficient of 2D BP reached 510
\si{\micro\volt/\kelvin} at 210 \si{\kelvin}, which is much higher
than the bulk BP (340 \si{\micro\volt/\kelvin} at 300
\si{\kelvin}).~\cite{saito2016gate} Monolayer BP also exhibits a
strong spatial anisotropy in electrical and thermal conductivities,
which makes the $ZT$ in the armchair direction larger than that in the
zigzag direction.~\cite{fei2014enhanced,bolin15-elph} However, it is
known that the 2D BP reacts strongly with chemical species in air and
thus the thermoelectric device can be quickly degrading.  As for TMDs
such as MoS$_2$, MoSe$_2$, WS$_2$ and WSe$_2$, these materials show
thickness-dependent thermoelectric properties and maximum PF of about
0.34 and 0.15 \si{\watt/\kelvin^2\m} for n-type monolayer MoSe$_2$ and
p-type MoS$_2$ monolayers, respectively, which are much higher than
those of bulk (0.02 and 0.03 \si{\watt/\kelvin^2\m} for bulk n-type
MoSe$_2$ and p-type MoS$_2$,
respectively).~\cite{wickramaratne2014electronic}

In the family of 2D semiconductors, the band structure of monolayer
group III chalcogenides such as InSe, GaSe, or GaS are rather unusual,
having combination of a flat band at the top of the valence band and a
parabolic band at the bottom of conduction band.  This feature leads
to appearance of a very sharp peak in the electronic density of states
(DOS) at the top of the valence band and an almost constant DOS at the
bottom of the conduction
band.~\cite{zolyomi2013band,zolyomi2014electrons} A recent report by
Geim's group has shown that the carrier mobility in few-layer InSe may
exceed 10$^3$ \si{\cm^2\volt^{-1}\s} at room
temperature.~\cite{bandurin2016high} In an earlier experiment, Rhyee
et al. showed that the bulk InSe crystal exhibits a low thermal
conductivity, $\kappa < 1.2$ W/mK, at room
temperature,~\cite{rhyee2009peierls} and the thermal conductivity
decreases with increasing temperature ($0.74$ W/mK at 705 K), giving
$ZT=1.48$.  By using constant relaxation time in Boltzmann transport
theory, Wickramaratne et al.,~\cite{darshana15-inse} showed
thickness-dependent thermoelectric properties of 2D group III
chalcogenides.  From these results, it seems that both the electrical
and thermal transport properties of InSe are beneficial for
thermoelectric performance and efficiency with both high PF and $ZT$.
We thus expect that InSe in its 2D form could be a good thermoelectric
material.


\begin{figure}[t!]
  \centering
  \includegraphics[clip,width=7.5cm]{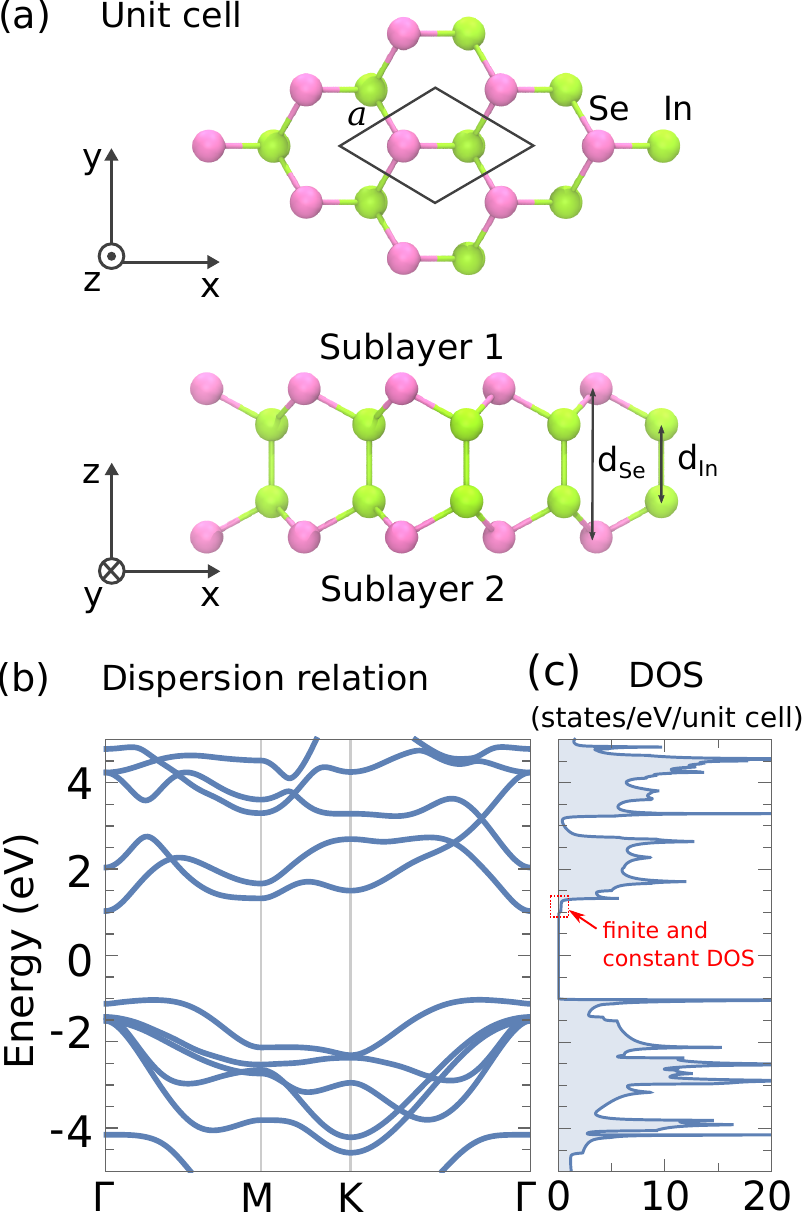}
  \caption{\label{fig:structure} Lattice structure and electronic
    properties of monolayer InSe.  (a) Top view and side view of the
    unit cell.  There are two sublayers in a monolayer InSe.  The $x$-
    and $y$-axes correspond to the armchair and zigzag directions,
    respectively.  (b) Electronic energy dispersion of monolayer InSe.
    (c) Density of states (DOS).  Relatively constant DOS in the
    conduction band is enclosed by a dotted box.}
\end{figure}

In this work, we evaluate the thermoelectric properties of monolayer
InSe by Boltzmann transport theory and first-principles calculations,
with a particular focus on its PF.  To calculate the constituents of
PF, i.e., $S$ and $\sigma$, we need the electronic energy dispersion
$E_{n\textbf{k}}$ and the carrier relaxation time $\tau_{n\textbf{k}}$
for each band $n$ and for each wave vector $\textbf{k}$.  The
electronic structure and relaxation time calculations are possible to
perform from first-principles by using \texttt{Quantum
  ESPRESSO}~\cite{giannozzi09-espresso} and electron-phonon Wannier
(EPW)~\cite{giustino07-epw,ponce16-epw} packages, respectively.  The
ground-state electronic structure is calculated within the
norm-conserving pseudopotential with the
Perdew-Zunger~\cite{perdew1981self} local density approximation
(LDA)~\cite{pseudoInSe} for the exchange-correlation functional and a
plane-wave basis set with kinetic energy cutoff of 160~Ry.  Note that
the LDA without spin-orbit interaction is adopted in this work because
the band gap is not really affected by the spin-orbit
interaction.~\cite{debbichi2015two} The system is modeled by adopting
a hexagonal supercell geometry where the vacuum distance is set to
12~\AA\ to eliminate the interactions between the InSe layer and focus
on the monolayer in the simulation.  To obtain the the optimized
geometry, the atomic positions and supercell vectors are fully relaxed
by using the Broyden-Fretcher-Goldfarb-Shanno minimization
method.~\cite{broyden1970convergence,fletcher1970new,goldfarb1970family,shanno1970conditioning}
This system is considered to be optimized when all the
Hellmann-Feynman forces and all components of the stress are less than
$5.0 \times 10^{-4}$ Ry/a.u. and $5.0 \times 10^{-2}$ GPa,
respectively, which are adequate for the present purpose.

The inputs for the EPW package are the electron energy and phonon
dispersions.  We compute the electron energy on a
$12\times 12\times 1$ $\textbf{k}$-point grid and the phonon
dispersion on a $6\times 6\times 1$ $\textbf{q}$-point grid.  The
electron-phonon matrix elements are first computed on these coarse
grids, and then interpolated to a dense mesh of both $\textbf{k}$ and
$\textbf{q}$-points of $150\times 150\times 1$ based on the maximally
localized Wannier functions in the EPW package because a dense finer
grid is necessary to evaluate transport
properties.~\cite{baroni01-dfpt} The main output of the EPW package is
the imaginary part of the electron self-energy
$\textrm{Im}(\Sigma_{n\textbf{k}}^{\textrm{e-ph}})$ calculated from
the interpolated ultra-dense electron-phonon matrix elements, which
then gives us the relaxation time $\tau_{n\textbf{k}}$ for each
electronic state by the relation
$({\tau_{n\textbf{k}}})^{-1} =
2[\textrm{Im}(\Sigma_{n\textbf{k}}^{\textrm{e-ph}})]/\hbar$,
where $\hbar$ is the reduced Planck constant. We use Gaussian
broadening with a small parameter of 10 meV to approximate the
$\delta$ function in
$\textrm{Im}(\Sigma_{n\textbf{k}}^{\textrm{e-ph}})$.  In the
calculation of scattering rate, the electron-phonon scattering due to
acoustic, optical, and polar optical phonons are all included within
the EPW package.~\cite{verdi2015frohlich,liu2017first} For simplicity,
in the calculation of the relaxation time ${\tau_{n\textbf{k}}}$ we do
not take into account the so-called ``momentum relaxation
time'',~\cite{lundstrom2009fundamentals} which is related to the
momentum loss in the scattering processes.

Having the information of the energy band structure and carrier
relaxation time, we can calculate the Seebeck coefficient $S$ and
electrical conductivity $\sigma$ along a certain direction ($x$- or
$y$-direction) by employing the Boltzmann transport theory within the
relaxation time approximation
(RTA):~\cite{goldsmid2010introduction,bolin15-elph,li2015electrical}
\begin{align}
  S &= -\frac{1}{eT} \frac{\displaystyle \sum_{n,\textbf{k}} (E_{n\textbf{k}} - E_F)
      v_{n\textbf{k}}^2 \tau_{n\textbf{k}} \frac{\partial
      f_{n\textbf{k}}}{\partial E_{n\textbf{k}}}}{\displaystyle \sum_{n,\textbf{k}} 
      v_{n\textbf{k}}^2 \tau_{n\textbf{k}} \frac{\partial
      f_{n\textbf{k}}}{\partial E_{n\textbf{k}}}}, \label{eq:S} \\
  \sigma &= - \frac{2e^2}{N V} \sum_{n,\textbf{k}} 
           v_{n\textbf{k}}^2 \tau_{n\textbf{k}} \frac{\partial
           f_{n\textbf{k}}}{\partial E_{n\textbf{k}}}, \label{eq:sigma}
\end{align}
where $e$ is the unit (positive) electric charge, $T$ is the average
temperature of the material, $N$ is the number of $\textbf{k}$ points,
$V$ is the volume of the unit cell using a constant thickness of 0.8
nm for the monolayer InSe,~\cite{bandurin2016high} $E_F$ is the Fermi
energy, $f_{n\textbf{k}}$ is the Fermi-Dirac distribution function,
and $v_{n\textbf{k}}$ is the component of the group velocity
($\nabla_{\textbf{k}} E_{n\textbf{k}}/\hbar$) in a particular
direction at each $\textbf{k}$ point, the factor 2 in
Eq.~\eqref{eq:sigma} accounts for the spin
degeneracy.~\cite{li2015electrical} Note that the RTA can usually be
justified for near-equilibrium transport and for specific types of
scattering (i.e., elastic, isotropic, or both), while in the case of
inelastic scattering (for example, by polar optical phonons which we
include in this study) we may have to consider additional
approximations.~\cite{kaasbjerg2012phonon} Nevertheless, the
relaxation time which includes the polar phonon scattering can still
be defined in EPW package from the imaginary part of the electronic
self-energy in the inelastic scattering.~\cite{kaasbjerg2012phonon}


Figure~\ref{fig:structure}(a) shows the top view and side view of the
unit cell of monolayer InSe with the lattice constant $a = 3.902$
\si{\angstrom}.  Two sublayers exists in a monolayer InSe, in which
the first and second sublayers are separated by
$d_{\text{In}} = 2.662$ \si{\angstrom} and $d_{\text{Se}} = 5.147$
\si{\angstrom} from the optimized geometry calculation.  In
Fig.~\ref{fig:structure}(b), we give the electronic structure of the
monolayer InSe from the LDA calculation.  The minimum point of the
first conduction band appears at the $\Gamma$ point, while the maximum
of the first valence band appears at a point along the $\Gamma$-$M$
direction.  The indirect band gap of monolayer InSe within the LDA is
about 2.06 eV.  Here $E_F = 0$ is set to be in the center of the
energy gap.  Since the Seebeck coefficients are sensitive to the
choice of the band gap, we also check the band gaps obtained by using
the Perdew-Burke-Ernzerhof (PBE)~\cite{perdew1996generalized} and the
Heyd-Scuseria-Ernzerhof (HSE)~\cite{zolyomi2014electrons} hybrid
functionals, which result in band gaps of about 1.55 eV and 2.24 eV,
respectively.  The HSE approach is the closest to the band gap of
monolayer InSe observed in the experiment.~\cite{bandurin2016high}
However, the EPW package does not support the HSE
pseudopotential. Therefore, the LDA with 2.06 eV band gap is a
reasonable approximation in this study.  Figure~\ref{fig:structure}(c)
shows the DOS of the monolayer InSe, in which we can see a very sharp
DOS at the top of the valence band.  In the conduction band, a finite
and almost constant 2D DOS appears for a limited range within 1.0--1.3
eV.  We argue that the existence of such DOS characteristics should be
relevant to the excellent thermoelectric of properties of the 2D
monolayer InSe that we will discuss below.

\begin{figure}[t]
  \centering
  \includegraphics[clip,width=8cm]{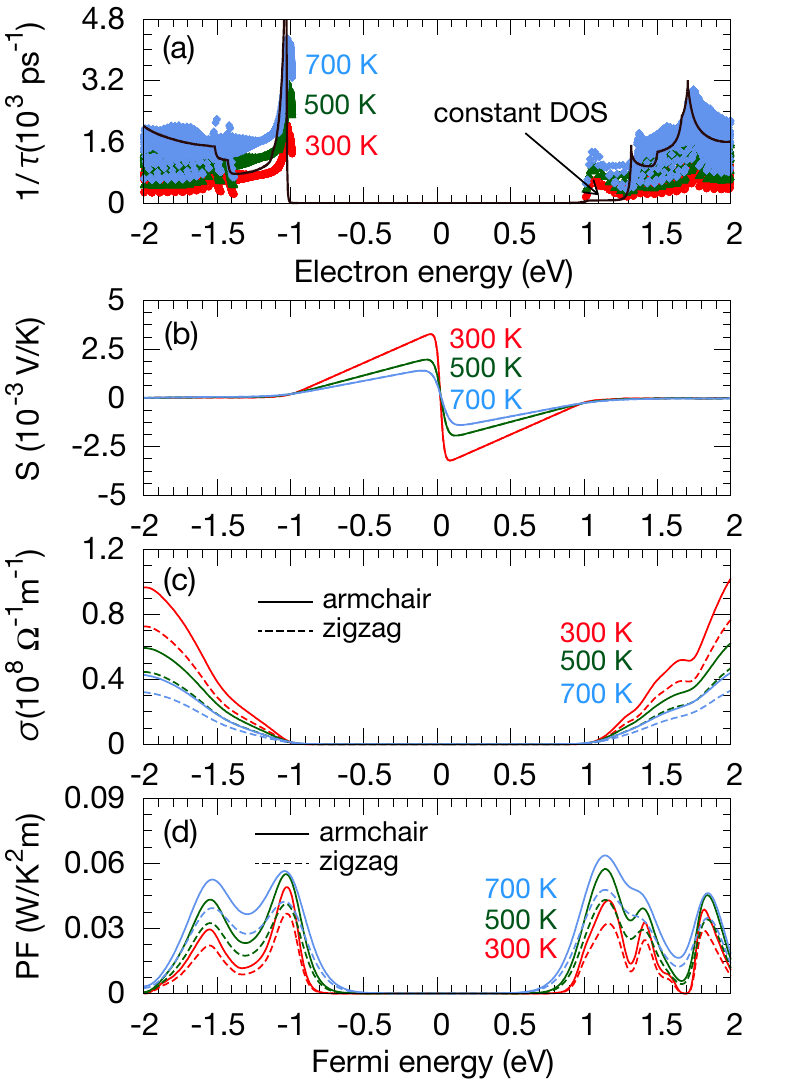}
  \caption{\label{fig:mainthermo} Transport and thermoelectric
    properties of monolayer InSe. (a) Carrier scattering rate (inverse
    of relaxation time) as a function of electron energy on a
    logarithmic scale.  Solid line is a schematic DOS of monolayer
    InSe in arbitrary units.  (b-d) Seebeck coefficient, electrical
    conductivity, and power factor as a function of Fermi energy for
    three different temperatures.  Solid and dashed lines in (b-d) for
    each temperature denote the quantities along the armchair and
    zigzag directions, respectively.  Note that the different
    directions give negligible effects for the Seebeck coefficients so
    that two curves for each temperature in (b) overlap each other.}
\end{figure}

In Figs.~\ref{fig:mainthermo}(a-d), we show the transport and
thermoelectric properties of monolayer InSe.  Firstly,
Fig.~\ref{fig:mainthermo}(a) depicts the scattering rate
($1/\tau_{n\textbf{k}}$) for all $\textbf{k}$ states in the electron
energy range $[-2.0,2.0]$ eV for three different temperatures: 300 K,
500 K, and 700 K at $E_F = 0$ eV.  Carrier densities needed to reach
$E_F = -2$ eV and $E_F = 2$ eV are $22\times 10^{21}$ cm$^{-3}$ and
$15\times 10^{21}$ cm$^{-3}$ for hole doping and electron doping,
respectively.  Since $1/\tau \propto DOS$, the shape of the $1/\tau$
curve resembles that of the DOS.  The increase of temperature also
enhances the relaxation time $\tau_{n\textbf{k}}$.  We find a
relatively small value of scattering rate at the conduction band of
monolayer InSe, leading to a larger relaxation time at energy around
1.0-1.3 eV. We thus expect that the power factor is enhanced in this
conducting regime for the n-type InSe. On the other hand, within the
valence band, the scattering rate is larger (the relaxation time is
smaller) but there is a very sharp DOS at energy around top of the
valence band which, according to the Mahan-Sofo
theory,~\cite{mahan1996best} is also a good region to obtain an
enhancement of the PF.

In Fig.~\ref{fig:mainthermo}(b), we show the Seebeck coefficient $S$
as a function of Fermi energy $E_F$. The rigid band approximation is
adopted to calculate the thermoelectric properties, which assumes that
the band structure remains unchanged as we move the Fermi level up and
down to simulate the electron and hole doping, respectively.  It is a
good approximation as long as the doping levels used are not high
enough to change the bonding properties of the
material.~\cite{goldsmid2010introduction} The larger $S$ is found at
lower temperature since $S \propto 1/T$ as given in Eq.~\eqref{eq:S}.
The maximum value of $S$ for monolayer InSe at room temperature can be
more than 3000 \si{\micro\volt/\kelvin}, which is also a very large
$S$ among the 2D materials.  This value of $S$ is mainly determined by
the band gap of the monolayer InSe, in which for band gaps much larger
than the thermal energy we can approximate the Seebeck coefficient to
be proportional to the band gap.~\cite{hung2015diameter}

The transport coefficients such as $S$ and $\sigma$ are often measured
in a particular direction.  In the case of monolayer InSe, we have
defined the armchair and zigzag direction from the $x$- and $y$-axes
as shown in Fig.~\ref{fig:structure}(a).  Therefore, besides the
temperature dependence, there is also an orientation dependence of the
transport coefficients.  However, for the Seebeck coefficient in
Fig.~\ref{fig:mainthermo}(b), we find that the values of the two
orientations almost overlap to each other.  The origin of this
behavior can be traced back to the contribution of the group velocity
in both the numerator and the denominator parts of Eq.~\eqref{eq:S},
where the contribution from group velocity might be vanished by the
division.  On the other hand, we show in Fig.~\ref{fig:mainthermo}(c)
that the electrical conductivity $\sigma$ depends on the crystal
orientation because only one group velocity term appears in the
expression of $\sigma$ in Eq.~\eqref{eq:sigma} and the band structure
(thus group velocity) of monolayer InSe is slightly anisotropic.  We
find that the armchair direction of monolayer InSe gives a larger
$\sigma$ than the zigzag direction.  For example, at room temperature
and $E_F \approx 1.15$ eV (the location of PF maximum in n-type
monolayer InSe), we have $\sigma = 0.056 \times 10^8$ S/m for the
armchair direction, while $\sigma = 0.042 \times 10^8$ S/m for the
zigzag direction.

By combining $S$ and $\sigma$, we can calculate PF $=S^2 \sigma$ as a
function of Fermi energy, as shown in Fig.~\ref{fig:mainthermo}(d).
The largest PF should be obtained for the armchair direction at
the high temperature in \emph{both} p-type and n-type monolayer InSe
(PF value of the p-type monolayer InSe is close to that of the n-type).
This result slightly differs with that calculated by Wickramaratne et
al.,~\cite{darshana15-inse} which stated that the p-type monolayer
InSe has a much larger PF than the n-type.  The discrepancy should
come from the fact that they treated the relaxation time $\tau$ as a
constant,~\cite{darshana15-inse} while in this work we consider $\tau$
to be energy-dependent as a result of taking the electron-phonon
scattering into account.  Nevertheless, we note that the PF of p-type
monolayer InSe is still on the same order of magnitude with that of
the n-type.  Therefore, experimentalists could have flexibility to
dope monolayer InSe and to obtain the most optimized PF depending on
the device setup.  From our calculation, the carrier density needed to
reach the maximum PF in the p-type (n-type) monolayer InSe is about
$2.4 \times 10^{21}$ ($0.11 \times 10^{21}$) cm$^{-3}$ by hole
(electron) doping.

The high PF in monolayer InSe originates from both the large $S$ and
$\sigma$, corresponding to the unique band structure of monolayer InSe
with semiconducting and unusual shape of DOS,
respectively. Furthermore, we have shown in a previous work that one
way to obtain large PF is by using a low-dimensional semiconductor
with high intrinsic carrier mobility and small confinement length $L$
(the thickness of the 2D material in this
case).~\cite{hung2016quantum} The confinement length $L$ of monolayer
InSe is found to be very small, about 0.8 nm, compared with thermal de
Broglie wavelength $\Lambda \sim 10$ nm at room
temperature~\cite{bandurin2016high} and thus improving the PF of the
monolayer InSe compared to its bulk form. Wickramaratne et
al.,~\cite{darshana15-inse} reported PF $=0.006$ ($0.001$) W/K$^2$m
for bulk p-type (n-type) InSe, which is much smaller than that of 2D
InSe with PF $=0.049$ ($0.043$) W/K$^2$m for monolayer p-type (or
n-type) InSe. It should be noted that, not only the dimensionality,
but the scattering mechanisms and DOS could also be important to
enhance the PF.~\cite{kim2009influence}

In conclusion, monolayer InSe has been shown to be a potential
thermoelectric material with high PF.  We expect that a further
examination on the thermal conductivity $\kappa$ could suggest us the
best value of $ZT$.  As a rough estimate, if we use $\kappa$ of
monolayer InSe of about 27
\si{\watt/\meter\kelvin},~\cite{nissimagoudar2017thermal} combined
with PF $=0.049$ ($0.043$) W/K$^2$m, it might be possible to achieve
$ZT$ of about 0.54 (0.48) for p-type (n-type) InSe at room
temperature.  Other group III chalcogenides materials such as GaS and
GaSe in their 2D forms, having similar band structures with InSe, may
also be good candidates for thermoelectric applications.

\section*{Acknowledgments}
N.T.H. and A.R.T.N are supported by the Interdepartmental Doctoral
Degree Program for Multidimensional Materials Science Leaders in
Tohoku University.  R.S. acknowledges JSPS KAKENHI Grant Numbers
JP25107005 and JP25286005.

%

\end{document}